\documentclass{ecai}

\usepackage{amsmath, amssymb, amsfonts, mathtools, amsthm}
\usepackage{algorithm, algorithmic}
\usepackage{pifont}

\usepackage{booktabs}
\usepackage{multirow}
\usepackage{array}
\usepackage{graphicx}
\usepackage{subcaption}
\usepackage{caption}
\usepackage{tabu}
\usepackage{float}
\usepackage{adjustbox}
\usepackage{siunitx}
\usepackage{color}
\usepackage[table]{xcolor}
\usepackage{colortbl}

\usepackage{microtype}
\usepackage{textcomp}
\usepackage{xspace}
\usepackage{enumitem}

\usepackage{latexsym}
\usepackage{cuted}
\usepackage{capt-of}
\usepackage{hyperref}
\hypersetup{hidelinks=true}
\usepackage[capitalize,noabbrev]{cleveref}
\usepackage{cite}
\usepackage[switch]{lineno}
\usepackage{todonotes}

\theoremstyle{plain}

\theoremstyle{definition}

\theoremstyle{remark}

\newcommand{\BibTeX}{B\kern-.05em{\sc i\kern-.025em b}\kern-.08em\TeX}

\newbool{showComments}
\booltrue{showComments}

\ifbool{showComments}{%

}

\newcommand{\SysName}{Q-HEART\xspace} 
\newcommand{\barcell}[2]{%
  \color{#2}\rule{#1pt}{8pt}%
}

\begin{document}

\begin{frontmatter}

\title{\SysName: ECG Question Answering via Knowledge-Informed Multimodal LLMs}

\author[A]{\fnms{Hung Manh}~\snm{Pham}}
\author[B]{\fnms{Jialu}~\snm{Tang}}
\author[B]{\fnms{Aaqib}~\snm{Saeed}} 
\author[A]{\fnms{Dong}~\snm{Ma}\thanks{Corresponding Author. Email: dongma@smu.edu.sg \\ Preprint}}

\address[A]{Singapore Management University}
\address[B]{Eindhoven University of Technology}

\begin{abstract} 
Electrocardiography (ECG) offers critical cardiovascular insights, such as identifying arrhythmias and myocardial ischemia, but enabling automated systems to answer complex clinical questions directly from ECG signals (ECG-QA) remains a significant challenge. Current approaches often lack robust multimodal reasoning capabilities or rely on generic architectures ill-suited for the nuances of physiological signals. We introduce \SysName, a novel multimodal framework designed to bridge this gap. \SysName leverages a powerful, adapted ECG encoder and integrates its representations with textual information via a specialized ECG-aware transformer-based mapping layer. Furthermore, \SysName leverages dynamic prompting and retrieval of relevant historical clinical reports to guide tuning the language model toward knowledge-aware ECG reasoning. Extensive evaluations on the benchmark ECG-QA dataset show \SysName achieves state-of-the-art performance, outperforming existing methods by a $4$\% improvement in exact match accuracy. Our work demonstrates the effectiveness of combining domain-specific architectural adaptations with knowledge-augmented LLM instruction tuning for complex physiological ECG analysis, paving the way for more capable and potentially interpretable clinical patient care systems. 

\end{abstract}
\end{frontmatter}

\section{Introduction}
\label{sec:intro}

\begin{figure*}[t]     
\centering
\includegraphics[width=\linewidth]{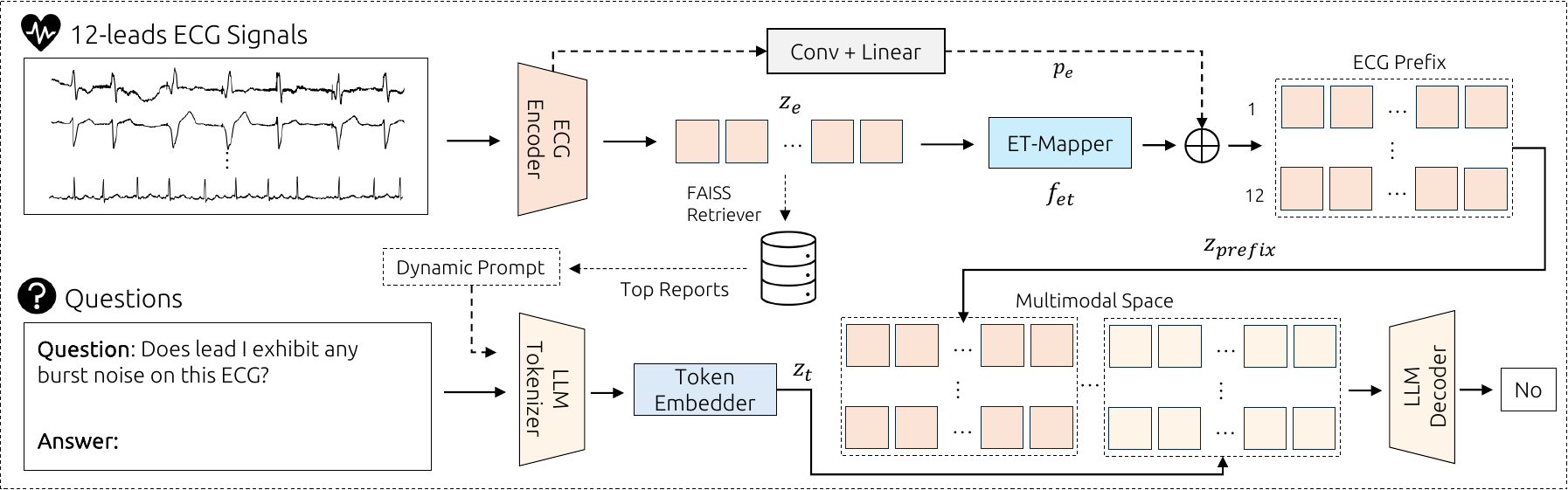}
\vspace{0.05cm}
\caption{Overview of our proposed \SysName framework.}
\label{fig:SysName} 
\vspace{0.6cm}
\end{figure*}

Electrocardiograms (ECGs) are non-invasive, cost-effective diagnostic tools that play a pivotal role in detecting cardiac arrhythmias and other cardiovascular abnormalities in clinical practice~\citep{zhang2023self,na2024guiding, liuzero, pham2024c}. A standard clinical ECG typically involves 12 leads, each capturing electrical activity from different anatomical perspectives of the heart, offering a comprehensive view of cardiac function. Recently, single or few-lead ECG acquisition has also become commonly used in wearables, facilitating large-scale, continuous monitoring in both clinical and consumer settings. As the volume and accessibility of ECG data increase, there is a pressing demand for intelligent systems that can interpret these data and assist clinicians or patients in understanding cardiac health. Accordingly, the need for advanced systems that can respond to clinical questions about ECGs has become more essential. ECG-based question answering (ECG-QA) aims to address this challenge by enabling automated systems to generate precise and clinically relevant answers to questions derived from ECG data.

In parallel with these developments, large language models (LLMs) have rapidly emerged as foundational tools in artificial intelligence, demonstrating impressive capabilities in text generation, understanding, and reasoning. By leveraging their vast parameter spaces and pretraining on diverse datasets, LLMs can encode a wealth of knowledge, enabling them to handle complex reasoning and produce human-like responses. Recent models such as Med-Gemini~\citep{saab2024capabilities}, BioMedGPT~\citep{luo2023biomedgpt}, and AMIE~\citep{o2024towards} have achieved state-of-the-art performance across a variety of medical tasks. While LLMs have primarily been applied to textual data, the integration of non-linguistic inputs (particularly information-rich physiological signals like ECGs) is an emerging research direction. This leads directly to the central question of our work: \textit{can LLMs be effectively adapted to ground their reasoning in ECG signals to accurately answer clinical questions in an ECG-QA setting?}

Despite the promise of LLMs in healthcare, their specific application to interactive ECG-QA remains nascent, facing distinct and underexplored challenges. Specifically, existing efforts have mainly focused on related applications such as ECG report generation~\citep{zhao2024ecg, wan2024meit, tang2025electrocardiogram} and retrieval-augmented diagnosis~\citep{yu2023zero}, which, although valuable, fall short of enabling fine-grained, interactive question answering. Early works like ECG-QA~\citep{NEURIPS2023_d0b67349} pioneered this direction by curating a benchmark dataset and corresponding baselines, but showing limitations or underperformance in the weak ECG encoder and LLM usage, hinder their reasoning capabilities. Recent approaches, such as ECG-LM~\citep{yangecg} and Med-Gemini~\citep{saab2024capabilities}, have advanced ECG-text alignment through multimodal architectures. Nonetheless, they often require specialized training data (e.g., personalized information or GPT-generated templates), depend heavily on large proprietary models, and do not fully leverage efficient instruction tuning, automatic retrieval augmentation, or ECG lead-specific knowledge integration to achieve better multimodal reasoning in clinical QA settings.  

In this paper, we introduce \SysName, a novel LLM-based multimodal framework designed to address the emerging task of ECG-based question answering (ECG-QA). \SysName integrates ECG signal representations with textual embeddings through a specially designed transformer mapping layer, enabling accurate and contextually relevant answers to clinical questions. By leveraging carefully crafted instruction tuning, supported by retrieval-based historical reports from a large-scale database with dynamic prompting techniques, our approach enhances multimodal alignment, reasoning capabilities, and shows potential interpretability in ECG-QA tasks. We evaluate \SysName on the benchmark dataset, achieving state-of-the-art performance in exact-match accuracy and significant improvements in natural language generation metrics. The contributions of this work can be summarized as follows: 

\begin{itemize}
    \item We present \SysName, a novel framework for ECG question answering (ECG-QA) that synergistically integrates adapted deep ECG encoding, specialized multimodal mapping, and knowledge-informed LLM instruction tuning. \SysName establishes a new state-of-the-art on the benchmark ECG-QA task.
    \item We design the ET-Mapper, a novel ECG-aware transformer mapping layer, specifically developed to effectively bridge structural ECG signal representations with the LLM's textual embedding space, noticeably enhancing multimodal alignment.
    \item We incorporate retrieval augmentation using historical clinical reports, sourced via efficient similarity search, to inject crucial contextual knowledge into the LLM's reasoning process for ECG-QA, notably without requiring patient-specific data or restrictive templates.
    \item We introduce dynamic prompting strategies for instruction tuning, including randomized answer option shuffling and selective use of retrieved reports, demonstrably improving model robustness and overall performance. 
\end{itemize}

\section{Related Works}
\label{sec:relatedworks}

\subsection{ECG Representation Learning}

Over the years, the development of ECG representation learning has played a crucial role in enabling a wide range of downstream tasks such as arrhythmia classification and clinical report generation. In the context of ECG-based question answering, learning robust and expressive signal representations is also essential, as it directly affects the model’s ability to reason over physiological patterns and provide accurate answers. In this section, we present prior approaches based on self-supervised learning (SSL), which have proven to be highly effective in medical domains where labeled data is often scarce or expensive to obtain. We focus on two primary streams of work: unimodal ECG representation learning and multimodal ECG-text learning.

Unimodal ECG representation learning focuses on extracting informative signal-level features directly from raw ECG waveforms, without relying on auxiliary modalities. One line of early SSL research approaches this by training autoencoder-based models to reconstruct masked or corrupted input signals, thereby encouraging the model to capture essential morphological patterns and temporal dependencies~\citep{zhang2023self,hu2023spatiotemporal,na2024guiding}. In contrast, another branch of SSL methods emphasizes contrastive objectives, aiming to maximize the agreement between augmented views of the same ECG instance. These approaches employ domain-specific augmentation strategies. For instance, \citep{gopal20213kg} introduces physiologically motivated transformations such as vectorcardiogram (VCG) projections to capture the three-dimensional spatiotemporal characteristics of cardiac activity. Additionally, ~\citep{oh2022lead} proposes to use contrastive-based pretraining, Contrastive Multi-Segment Coding (CMSC) \citep{kiyasseh2021clocs}, and random lead masking to simulate diverse local and global lead configurations.

With the rise of multimodal learning, recent attempts have focused on aligning ECG signals with textual data, enabling zero-shot interaction across modalities for tasks like cardiac condition classification or ECG-report retrieval. Several works adopt contrastive learning frameworks such as ETP~\citep{liu2024etp}, MERL~\citep{liuzero}, ESI~\citep{yu2024ecg}, and others~\citep{lalam2023ecg}, with common architectures like ResNet or ConvNeXt for ECG and BERT-based models for text encoding, which demonstrate effective cross-modal alignment. More recently, C-MELT~\citep{pham2024c} introduces a hybrid self-supervised approach that hybridly combines contrastive learning with masked autoencoding, achieving state-of-the-art performance across a range of downstream ECG diagnosis tasks.

\subsection{ECG Report Generation}
Although our focus is on ECG-based question-answering, related works in text generation, such as diagnosis reports, share some similar goals of interpreting ECG data for clinical insights. Accordingly, several recent approaches have emerged in this domain. For example, \citep{bleich2024automated} proposes an encoder-decoder framework for ECG report generation, where ECG signals are processed using a ResNet encoder and decoded via LSTM or Transformer to produce clinical interpretations. 
\citep{qiu2023automated} further uses a multimodal ECG interpretation framework that encodes ECG signals as images and aligns them with diagnostic reports using a vision-language learning paradigm. Next, a recent work, ECG-Chat \citep{zhao2024ecg} introduces a multimodal framework leveraging contrastive learning to align ECG waveforms with diagnostic reports. This model is supported by a large instruction dataset and an innovative LaTeX report pipeline to improve user clarity and usability. Similarly, MEIT \citep{wan2024meit} proposes large language instruction-tuned models to address the complexities of ECG report generation. By introducing an attention-based fusion mechanism for ECG and text, MEIT achieves effective representational alignment between ECG signals and text. Finally, \citep{tang2025electrocardiogram} presents ECG-ReGen, a retrieval-augmented framework that combines ECG encoders with LLMs for zero-shot report generation.

\subsection{ECG Question Answering}

ECG-QA is a growing field that offers meaningful benefits, such as improving clinical decision-making and enhancing interactive patient care. Although it is still in its early stages, several studies have demonstrated promising results. For example, ECG-QA work \citep{NEURIPS2023_d0b67349} is the first significant contribution to this domain, comprising a novel dataset with ECG recordings, questions, and answers across diverse types, such as single-verify, single-choose, and single-query formats. The study also provides benchmarking results using various models in a classification manner (e.g., leveraging M$^3$AE \citep{chen2022multi}, SE-WRN\citep{han2021towards}, and common OpenAI model APIs), highlighting the potential for advancing ECG question-answering research. Following this, ECG-LM \citep{yangecg} also built upon the ECG-QA dataset by developing a multimodal large language model that integrates textual and ECG signal data. The framework employs a ResNET-18-based ECG encoder \citep{jing2021ecg} and usage of BioMedGPT-LM-7B \citep{luo2023biomedgpt} for text generation given detailed prompts, which overall shows better performance than the previous ECG-QA baselines. However, it relies on rich patient-related metadata (e.g., age, gender, weight, height) and exact clinical diagnosis reports from current examined datasets. 

Further, Med-Gemini \citep{saab2024capabilities} presents a comprehensive multimodal large language model capable of handling diverse medical tasks, including ECG-based QA. Although Med-Gemini shows state-of-the-art performance with its proposed instructed Gemini model, its methodology remains vague, especially for ECG embedding processing and ECG-QA instruction tuning, with limited implementation details disclosed. More recent works, like \citep{wang2025ecg, tang2024electrocardiogram} show their interesting ideas on ECG question-answering developments, such as a novel diverse dataset for evaluating diagnostic capabilities in ECG interpretation or a multimodal meta-learning framework for few-shot ECG QA, although they focus on their specific tasks and evaluations. 

Despite progress on the task, existing works face key limitations. First, they often rely on relatively weak ECG encoders, limiting physiological generalization. Second, the existing ECG-QA models lack development with ECG-aware knowledge in aligning ECG and text with a mapping architecture, which could strengthen multimodal ECG-text learning and decoding. Lastly, current approaches remain largely unexplored in the available usages of historical reports and textual diagnoses in LLM instruction tuning, leaving gaps in interpretability and critical reasoning capabilities for clinical trust.
\vspace{-0.1in}

\section{Methods}
\label{sec:methods}

Our approach, \SysName, introduces a multimodal framework specifically designed to enable LLMs to perform complex question-answering directly from ECG signals.  As illustrated in Figure~\ref{fig:SysName}, the core idea is to bridge the gap between physiological signal processing and natural language reasoning. First, \SysName processes the input ECG using a powerful, adapted deep encoder (Section~\ref{sub:ecg_encoder}) to capture condensed and informative ECG features. Simultaneously, to provide essential clinical context, relevant historical reports associated with similar ECGs are retrieved from a large-scale database (also Section~\ref{sub:ecg_encoder}). The crucial step of multimodal integration is handled by our proposed ET-Mapper (Section~\ref{sub:ET-Mapper}), an ECG-aware transformer architecture that transforms the ECG embeddings into a format compatible with the LLM's textual input space. These aligned ECG representations, along with the retrieved reports and the user's question, are then formulated into a dynamic instruction prompt (Section~\ref{sub:IT}). Finally, an instruction-tuned LLM decoder processes this fused input to generate a contextually relevant and accurate answer. The following sections delve into each key component of our model.

\subsection{ECG Encoder and Report Retrieval} \label{sub:ecg_encoder}

The ECG encoder in our model (denoted as $f_e$) is adapted from C-MELT's ECG encoder architecture~\citep{pham2024c}, designed for robust cross-modal ECG-text representation learning, which has been shown to perform notably well on various ECG downstream tasks, covering more than 100 cardiac conditions. Their ECG encoder employs multiple convolutional layers with GELU activation and group normalization for efficient initial feature extraction. These are followed by positional encoding and 8 transformer layers, enabling robust sequential learning of ECG signals. Ultimately, the features are projected into a $d=768$ dimensional space for efficient downstream tasks. Specifically for ECG signals, their encoder utilizes masking strategies such as random lead masking and segment dropout. Overall, their framework is enriched by contrastive modeling tasks in a multimodal learning manner using SigLIP-based loss~\citep{zhai2023sigmoid} and search-based negative sampling. In our pretraining step, we adopt and train the same design but extend it to 12 transformer layers. We also incorporate random lead masking with a high probability ($p = 0.65$) of masking each lead, motivated by the need that many questions pertain to a single or a few leads, encouraging the model to focus more on learning arbitrary lead-specific representations useful for downstream tasks. Ultimately, the trained ECG encoder is incorporated into the \SysName framework as in Figure~\ref{fig:SysName}. We present the performance of the adapted ECG encoder (referred to as C-MELTv2) in Section~\ref{subsub:validate_ecg_encoder}.

\begin{figure}[t]     
\centering\includegraphics[width=\columnwidth]{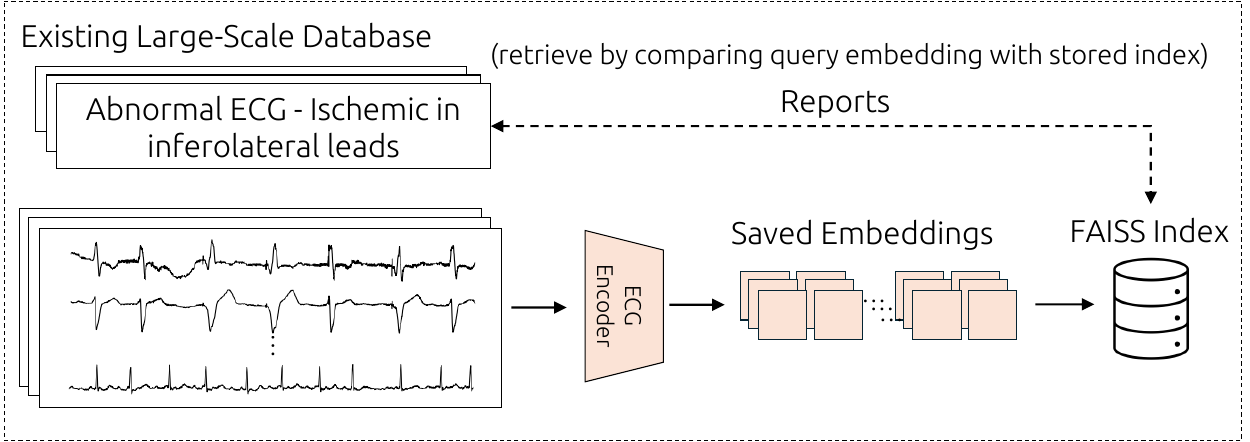}
    \caption{FAISS index construction in \SysName for efficient similarity search, it leverages large-scale ECG-report datasets.}
    \vspace{0.1cm}
    \label{fig:faiss}
    \vspace{0.8cm}
\end{figure}

To enable flexible clinical context and interpretability in the ECG-QA task, we propose an automatic information retrieval-based augmentation mechanism that consists of two stages: index construction and query-time retrieval. During the index construction phase, we leverage a large-scale ECG-report paired dataset such as MIMIC-IV ECG~\citep{gow2023mimic}. Each ECG signal in the dataset is encoded using our ECG encoder $f_e$, and the resulting embeddings are stored in a FAISS index~\citep{douze2024faiss}, with each entry linked to its corresponding clinical report (Figure~\ref{fig:faiss}). This enables scalable similarity search across a vast corpus of ECG data. At query time, the input ECG $\mathbf{x}_e$ is encoded into a dense vector $\mathbf{z}_e = f_e(x_e) \in \mathbb{R}^d$, which is then used to retrieve the top-3 most similar embeddings based on cosine similarity. The textual reports associated with these retrieved entries (denoted $\mathcal{R}_3$) are returned and incorporated into the dynamic prompting strategy. This design not only strengthens the contextual grounding of the LLM but also introduces external clinical knowledge that enhances answer accuracy and model interpretability.

\subsection{ET Mapper} \label{sub:ET-Mapper}

To bridge the modality gap between ECG signals and textual data, we introduce a transformer-based mapping called ET-Mapper. This component is responsible for transforming the $\mathbf{z}_e$ into a prefix ECG embedding that is compatible with LLM token embeddings $\mathbf{z}_t$. Formally, given an ECG embedding $\mathbf{z}_e$, the ET-Mapper learns a mapping function $f_{\text{et}}: \mathbb{R}^{d} \rightarrow \mathbb{R}^{c \times d'}$ that projects $\mathbf{z}_e$ into a sequence of $c=12$ channel embeddings, each of dimension $d'$ (dimension of LLM token embeddings). The choice of $c=12$ directly corresponds to the 12 standard leads in a clinical ECG, which aims to produce the prefix tokens that structurally reflect the inherent spatial diversity of ECG signals. More specifically, the transformation begins by projecting $\mathbf{z}_e$ through a fully connected linear layer, which reshapes it into an initial prefix sequence of shape $\mathbb{R}^{c \times d'}$. This sequence is then refined by a lightweight transformer architecture comprising $L=2$ layers with $H=4$ attention heads. The transformer enables the deeper modeling of ECG features within the prefix sequence, allowing the final output $\mathbf{z}_{et} = f_{\text{et}}(\mathbf{z}_e)$ to serve ECG-text alignment during LLM-based decoding. 

However, at this stage, the lead positional information from individual ECG leads turns out to be unclear (due to the nature of the global characteristics from $\mathbf{z}_e$). Since clinical questions might target specific leads of the ECG signal, it is essential to retain and inject lead-aware spatial structure into the prefix representation. To address this, we incorporate an auxiliary skip connection from the ECG encoder's positional embedding layer into the ET-Mapper. Specifically, we extract the positional lead representation from the encoder (denoted as $\mathbf{p}_e \in \mathbb{R}^{c \times d'}$) via a lightweight convolutional layer followed by a linear projection. The result $\mathbf{p}_e$ is then added element-wise to $\mathbf{z}_{et}$, and the final mapped prefix embedding is updated as $\mathbf{z}_{prefix} = \mathbf{z}_{et} + \mathbf{p}_e$. This additive fusion enriches the ECG prefix embedding with localized lead-awareness, ensuring that the language model receives input conditioned not only on the overall ECG context but also on structurally lead-specific cues.

\subsection{Instruction Tuning} \label{sub:IT}
\begin{figure}[t]     
\centering\includegraphics[width=\linewidth]{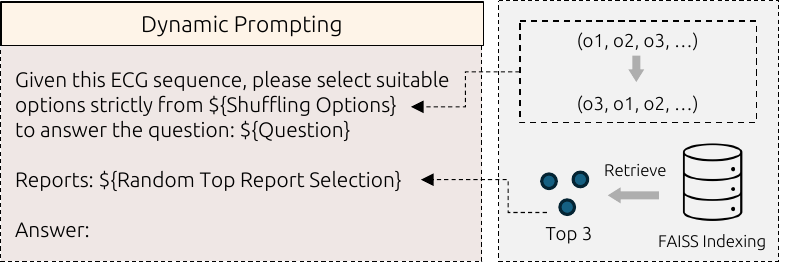}
    \caption{Dynamic prompting to guide instruction tuning in \SysName framework. This particularly includes shuffling options and random top report selection.}
    \label{fig:prompt}
    \vspace{0.8cm}
\end{figure}

To empower \SysName with relevant reasoning and strong clinical text generation capabilities, we adopt instruction tuning on a pretrained large language model, specifically leveraging the LLaMA-Instruct 1B~\citep{llama3instruct}. Despite its compact size, LLaMA-Instruct offers robust general-purpose instruction-following skills, making it a suitable foundation for our resource-efficient setup. Our tuning specifically follows a simple yet effective prompting strategy as illustrated in Figure~\ref{fig:prompt}, with an emphasis on two dynamic components that vary across training steps (denoted $DP$): (1) To prevent position bias and encourage robust understanding of candidate options, we randomly shuffle the order of multiple-choice options ($o$) during training. This ensures the model learns to focus on the semantic content rather than memorizing positional patterns. We present our strategies to specify the possible options for the task in Table~\ref{tab:options}. (2) We randomly select one of the top-3 retrieved reports $\mathcal{R}_3$ (as described in Section~\ref{sub:ecg_encoder}), at every training step. This promotes diverse evidence grounding and prevents overfitting to fixed report-query pairs.

Technically, during the forward pass, the prompt-formatted question is first tokenized into a sequence $\mathbf{x}_t$, which is then passed through the LLM’s token embedding layer $f_t$ to obtain token embeddings $\mathbf{z}_t = f_t(\mathbf{x}_t) \in \mathbb{R}^{l \times d'}$, where $l$ denotes the number of tokens. Subsequently, we concatenate those text embeddings with previously-defined ECG prefix as a single embedding $\mathbf{z}_{fuse} = concat(\mathbf{z}_{prefix}, \mathbf{z}_t) \in \mathbb{R}^{ (c+l) \times d'}$. The fused embedding is subsequently fed into the LLM decoder to perform the task of answer generation.
Here, we fine-tune instead of freezing LLaMA-Instruct 1B, as it naturally exhibits general-context reasoning capabilities but might lack rich domain-specific grounding in biomedical scenarios such as ECG reports and cardiology-related QA tasks. Here,  we use LoRA~\citep{hu2021lora} method with $r=8$, $\alpha = 32$, and dropout = 0.1 settings, while employing standard cross-entropy loss in autoregressive language modeling. 

\begin{table*}[t]
\centering
\caption{Construction rules for candidate answer options based on question type and content. These options are appended to the prompt during instruction tuning. Note that <attribute list> is provided in the dataset itself and also used by other existing methods.}
\label{tab:options}
\begin{tabular}{@{}llp{8cm}@{}}
\toprule
\textbf{Question Type} & \textbf{Condition} & \textbf{Candidate Options} \\
\midrule
Single-Query & Starts with "What leads" & lead I, lead II, lead III, lead aVR, lead aVL, lead aVF, lead V1, lead V2, lead V3, lead V4, lead V5, lead V6 \\
             & Starts with "What numeric features" & rr interval, p duration, pr interval, qrs duration, qt interval, qt corrected \\
             & Starts with "What range" & below the normal range, within the normal range, above the normal range \\
             & Other cases & <attribute list>, none \\
\midrule
Single-Verify & Any & yes, no, not sure \\

\midrule
Single-Choose & Any & <attribute list>, none \\
\bottomrule
\end{tabular}
\end{table*}

\section{Experiments}
\label{sec:exp}

\subsection{Datasets}
In our study, we use four public datasets, including (1) MIMIC-IV-ECG v1.0~\citep{gow2023mimic}, PTB-XL~\citep{wagner2020ptb}, (3) CODE-15~\citep{ribeiro2020automatic}, and (4) ECG-QA~\citep{NEURIPS2023_d0b67349}. Here, ECG-QA is the main dataset to fine-tune our proposed model on the question-answering task as the main experiment. Below, we briefly describe each dataset and its specific role in our framework.

\textit{MIMIC-IV-ECG v1.0}. This includes 800,035 paired samples from 161,352 unique subjects with numerous 10-second ECG recordings sampled at 500 Hz and the corresponding text reports. We use this large-scale dataset to pretrain ECG-text multimodal model and build FAISS indexing construction, which eventually aims for an effective ECG encoder and contextual report retrieval, respectively, for the ECG-QA task. 

\textit{PTB-XL}. The PTB-XL dataset includes 21,837 ECG signals collected from 18,885 patients. Each sample has a 12-lead ECG recording sampled at 500 Hz over 10 seconds and corresponding cardiac labels. Particularly, this dataset is commonly split into four sub-datasets (super - 5 classes, sub - 23 classses, form - 19 classes, and rhythm - 12 classes). Following the prior works~\citep{pham2024c, liuzero, wagner2020ptb}, we have processed the test set of each group (2158, 2158, 880, 2098 samples, respectively), which can be used to validate the zero-shot performance of the trained ECG encoder.

\textit{CODE-15}. We also validate the trained ECG encoder by using the test dataset of the CODE-15 dataset. Specifically, it contains 827 12-lead ECG samples of various lengths and a sampling rate of 400 Hz, covering 6 classes. Here, we ensure the processed signals are resampled at 500 Hz and have a length of 10 seconds. 

\begin{figure*}[t]     
\centering\includegraphics[width=\linewidth]{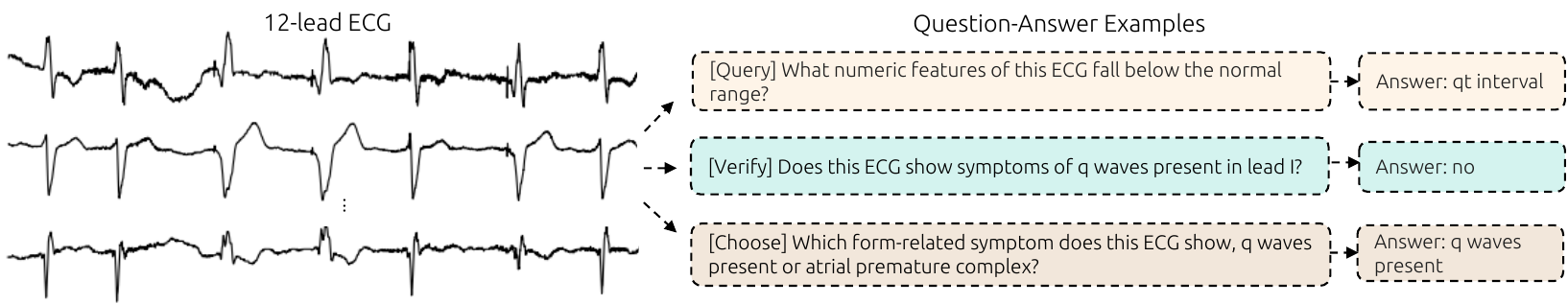}
    \caption{Examples of three question-answer pairs for a given 12-lead ECG signal.}
    \label{fig:data-example}
    \vspace{0.5cm}
\end{figure*}

\textit{ECG-QA}. Finally, this specialized dataset comprises curated questions and corresponding answers about various crucial cardiac aspects on the PTB-XL ECG samples. We focus on single-verify, choose, and query question types, yielding 159306, 31137, and 41093 samples for train, valid, and test sets. We present a more detailed description of this dataset as shown in Table~\ref{tab:data_info}. Particularly, we notice that many questions contains the word "lead" (approximately 60\%), which shows that this ECG-QA dataset also focuses on medical patterns of certain leads, rather than just giving attention to overall 12-lead ECGs as global diagnoses. This motivates us to put more care into the ECG positional encoding, as designed in Section~\ref{sub:ET-Mapper}. Finally, we also provide an example of an ECG sample with its corresponding questions and answers in Figure~\ref{fig:data-example}.

\begin{table}[t]
\centering
\caption{ECG-QA data description across three subsets (train, valid, test). For each set, we show the number of samples for each question type and, and at the end, the percentage of questions that have 'lead' care out of the total number of questions.}
\vspace{0.1cm}

\label{tab:data_info}
\begin{tabular}{l c c c l}
\toprule
\textbf{Set} & \textbf{Query} & \textbf{Verify} & \textbf{Choose} & \textbf{`Lead' Question (\%)} \\
\midrule
Train & 46737 & 62554 & 50015 & \barcell{65}{orange}~60\% \\
Valid & 11334  & 10718 & 9085 & \barcell{50}{teal}~57\% \\
Test  & 18157  & 13081 & 9855 & \barcell{55}{brown}~58\% \\
\bottomrule
\end{tabular}
\end{table}

\begin{figure*}[t]
\centering
\begin{minipage}{0.49\linewidth}
    \centering
    \includegraphics[width=\linewidth]{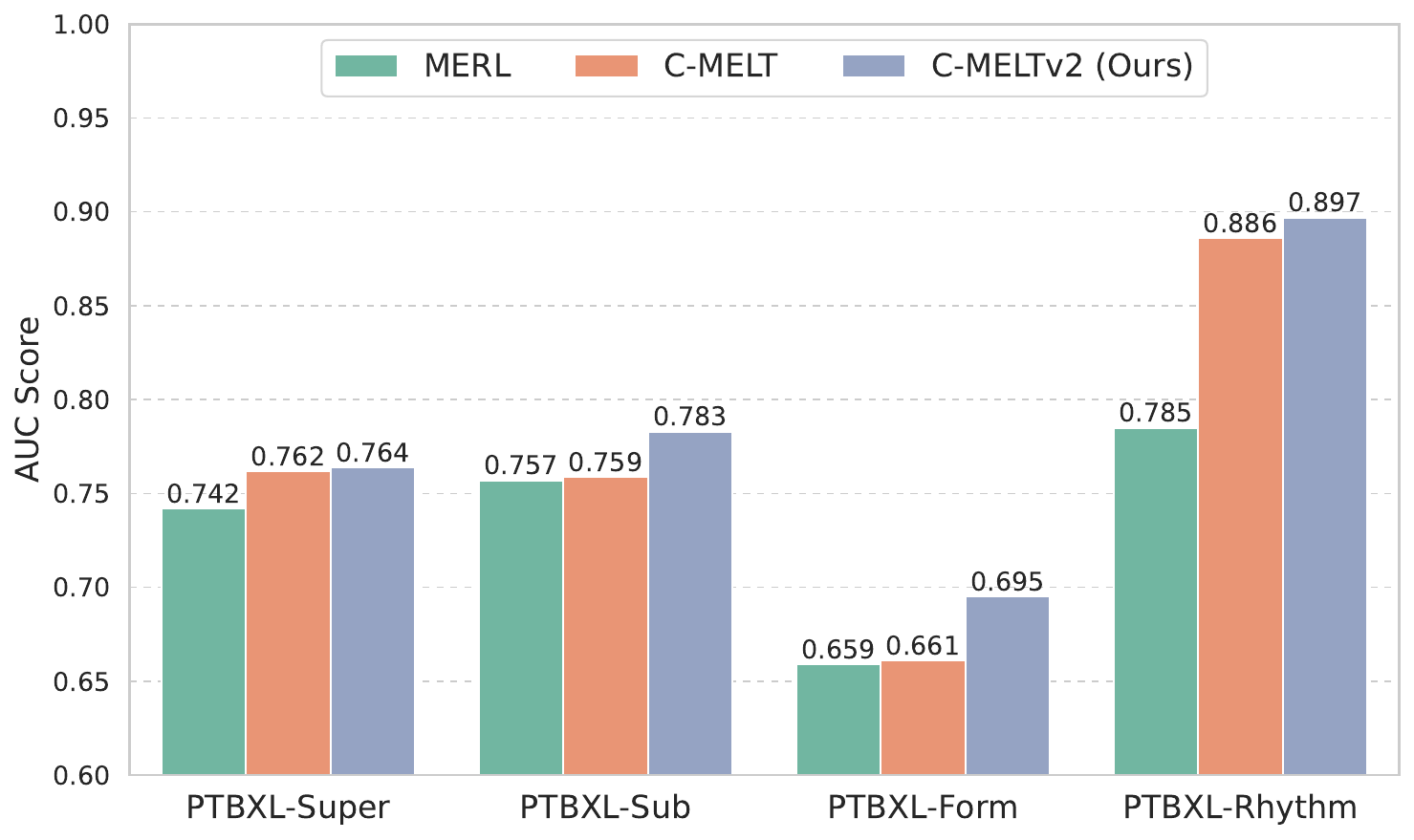}
    \caption{Zero-shot performance of different ECG encoders on the PTB-XL dataset.}
    \label{fig:ptbxl}
\end{minipage}
\hspace{0.01\linewidth} 
\begin{minipage}{0.49\linewidth}
    \centering
    \includegraphics[width=\linewidth]{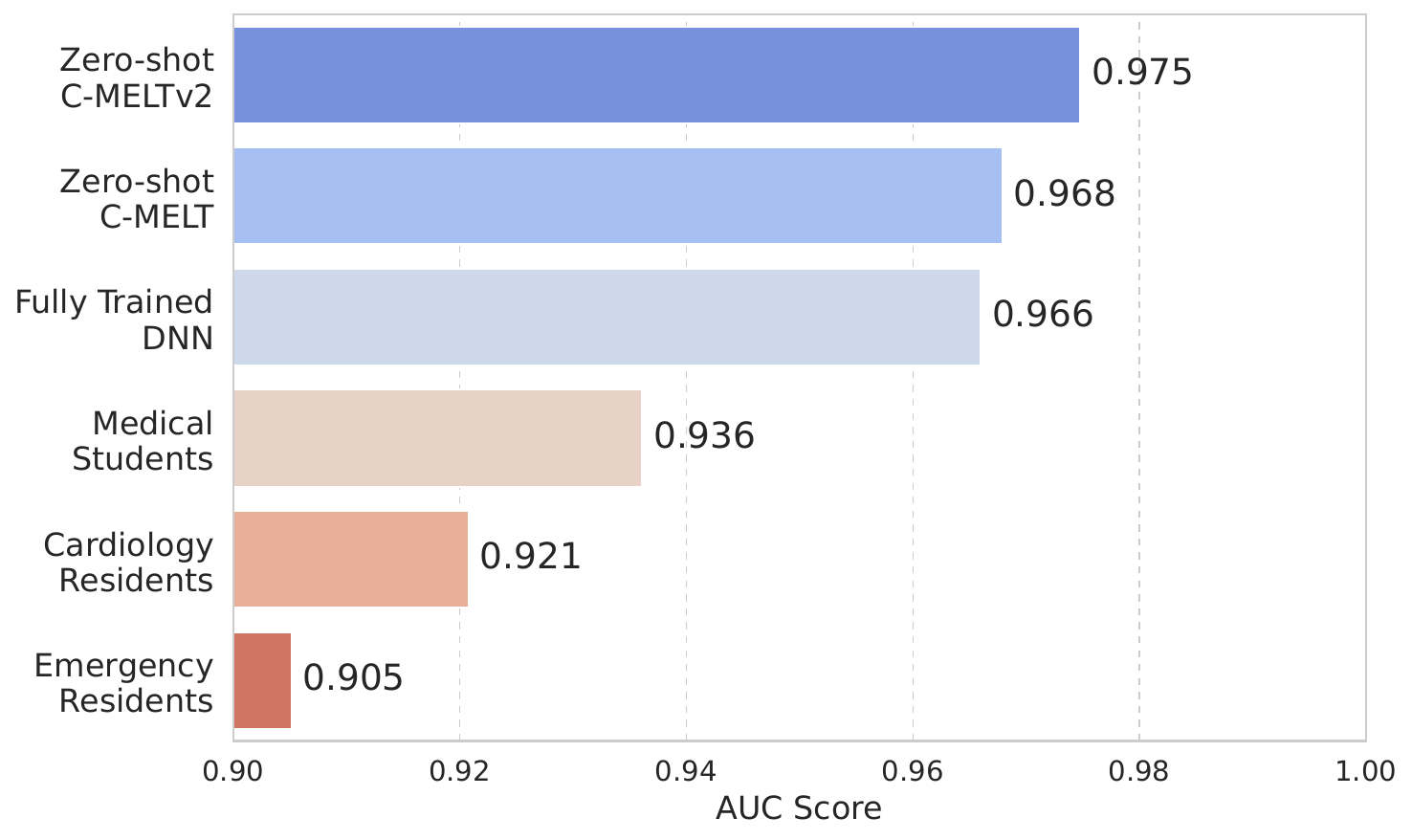}
    \caption{Performance of human experts vs. C-MELTv2 (ours) and other ECG encoders on the CODE-15 dataset.}
    \label{fig:test15}
\end{minipage}
\vspace{0.5cm}
\end{figure*}

\subsection{Training Details}

For the ECG encoder pretraining, we follow the training configurations in C-MELT report. For \SysName in detail, the model is fine-tuned using the AdamW optimizer with a learning rate of $5 \times 10^{-5}$, employing a cosine learning rate scheduler with a 0.1 warmup ratio. The optimizer is configured with $\beta_1 = 0.9$, $\beta_2 = 0.98$, an epsilon value of $1 \times 10^{-6}$, and a weight decay of 0.01. The model is trained for only 5 epochs with a batch size of 32 on a single NVIDIA H100-80GB GPU. During training, the best-performing model corresponding to the lowest validation loss is saved for later evaluation.

\subsection{Baselines and Metrics}

\subsubsection{Baselines}

To comprehensively evaluate the performance of our proposed approach, we benchmark \SysName against a diverse set of baseline models spanning multiple design paradigms.

First, for validating the trained ECG encoder, we compare it with two state-of-the-art ECG encoders, \textit{i.e.}, C-MELT~\citep{pham2024c} and MERL~\citep{liuzero},  which are well-pretrained on large-scale ECG-text datasets and have demonstrated strong performance across a variety of downstream cardiovascular tasks. These comparisons help isolate the contribution of our ECG representation module and justify its integration within our broader pipeline.

For the main task of ECG-QA, the evaluation is carried out under two regimes: using the entire testing data of ECG-QA dataset, and using only 10\% of it to assess robustness in low-resource settings~\citep{NEURIPS2023_d0b67349}. More importantly, we note that the baselines differ in how they utilize large language models, and we categorize them into four groups accordingly. (1) \textit{API}: These methods leverage off-the-shelf LLMs, including SE-WRN~\citep{zhong2019squeeze} paired with GPT-4, GPT-3.5-turbo, and text-davinci-003~\citep{NEURIPS2023_d0b67349} and ECG-ReGen~\citep{tang2025electrocardiogram}, where ECG signals are first converted into descriptive prompts and then fed into LLMs for answer prediction. (2) \textit{None}: The methods do not employ LLMs at all and instead form the ECG-QA task as a classification task following strategies introduced in~\citep{NEURIPS2023_d0b67349}, such as M$^3$AE~\citep{chen2022multi}, MedViLL~\citep{moon2022multi}, the Fusion Transformer, the Blind Transformer (text-only), and the Deaf Transformer (ECG-only). (3) \textit{Frozen}: Med-Gemini~\citep{saab2024capabilities} and ECG-LM~\citep{yangecg}, which align ECG signals and textual inputs within a shared space while retaining the original LLM parameters. (4) \textit{Tuning}: including our proposed \SysName and also Med-Gemini in its fine-tuned variant, adapting LLM parameters via instruction tuning to improve performance. 

\subsubsection{Evaluation Metrics}
In our work, we use six primary metrics for evaluation: Macro AUC score (0-1), Exact-Match Accuracy (EM-Acc, 0-1), BERTScore~\citep{zhang2019bertscore}, ROUGE~\citep{lin2004rouge}, METEOR~\citep{denkowski-lavie-2011-meteor}, and BLEU-1~\citep{papineni2002bleu}. Here, we use the macro AUC score in validating ECG encoder effectiveness, as followed by prior works~\citep{pham2024c, liuzero}, while EM-Acc serves as the main performance metric for comparing our model with existing works~\citep{NEURIPS2023_d0b67349, yangecg, saab2024capabilities}, as they only report this standard metric with strict correctness of the generated answers by comparing to the ground truth. However, EM-Acc alone may not fully capture the model’s reasoning quality or partial correctness, especially in complex query-type questions. For instance, a response that conveys the correct diagnosis using slightly different phrasing would still be penalized under EM-Acc, despite being clinically valid. Therefore, we complement this metric with natural language generation (NLG) metrics (BERTScore, ROUGE, METEOR, and BLEU-1) which offer deeper insight into the semantic alignment, lexical overlap, and fluency of generated answers. We use these metrics in our ablation studies, where finer-grained differences in model outputs need to be interpreted beyond binary correctness.

\subsection{Quantitative Results}

\subsubsection{ECG Encoder Validation} \label{subsub:validate_ecg_encoder}

It is essential to have a strong ECG encoder to extract informative representations for the downstream task of multimodal ECG-based question answering. In this section, we conduct a comprehensive validation of our encoder’s representational quality. As described in section~\ref{sub:ecg_encoder}, we trained entirely from scratch without relying on any pretrained weights, so this necessitates a verification of the downstream performance. Specifically, we design zero-shot 
classification experiments, following~\citep{pham2024c} on the CODE-15 and PTB-XL datasets to test generalization without any fine-tuning. While evaluating on the PTB-XL dataset is important because it represents the signal data distribution that is used in the ECG-QA dataset, CODE-15 also provides useful insights as it allows us to compare the performance between models and human experts. 

As can be seen in Figure~\ref{fig:ptbxl}, our trained ECG encoder (C-MELTv2) archieves the best performance, averaging of 78.5\% (AUC) over four PTB-XL sub-datasets, higher than the original C-MELT and MERL by nearly 2\% and 5\%, respectively. These improvements not only validate the original design from C-MELT but also benefit from enhancements such as a stronger transformer decoder and lead masking strategies. C-MELTv2 also shows its superior performance in the CODE-15 dataset evaluation, as shown in Figure~\ref{fig:test15}. Specifically, its zero-shot result is slightly better than the original C-MELT's and fully trained DNN model~\citep{ribeiro2020automatic}, but still show clear improvement (4\%-7\%) over human experts like medical students or in-domain residents. This further confirms the practical clinical relevance of our encoder in assisting or augmenting expert decision-making in real-world scenarios.
 
\subsubsection{ECG Question Answering Performance}

\begin{table*}[t]
\centering
\caption{Comparison of different LLM-based approaches on three question types: Verify, Choose, and Query (report in exact-match accuracy using the same test data configuration of the ECG-QA dataset).}
\vspace{0.05cm}

\label{tab:accuracy_comparison}
\setlength\tabcolsep{8pt}
\begin{tabular}{lcccccc}
\toprule
\textbf{Method} & \textbf{Test Data} & \textbf{LLM Usage} & \textbf{Verify} & \textbf{Choose} & \textbf{Query} & \textbf{Avg} \\
\midrule
SE-WRN+gpt-4\citep{NEURIPS2023_d0b67349} & 10\%  & API & 0.710& 0.481& 0.357& 0.516 \\
SE-WRN+gpt-3.5-turbo\citep{NEURIPS2023_d0b67349}  &  10\%& API&  0.693 & 0.361 & 0.311 & 0.455 \\
SE-WRN+text-davinci-003\citep{NEURIPS2023_d0b67349}  &  10\%& API & 0.750 & 0.378 & 0.360 & 0.496 \\
M$^3$AE\citep{chen2022multi} & 10\%& None & 0.760 & 0.582 &   0.400 & 0.581 \\
ECG-LM\citep{yangecg}  &  10\% & Frozen& 0.758 & 0.574 & 0.399 & 0.577 \\
Med-Gemini\citep{saab2024capabilities}  &  10\% & Frozen& -- & -- & -- & 0.577\\
\rowcolor{green!5}
\SysName (Ours)  &  10\%& Tuning & 0.913 & 0.626 & 0.345 & 0.628 \\
\midrule
ECG-ReGen\citep{tang2025electrocardiogram}   & All& API &  0.725 & 0.585 & 0.326 & 0.545 \\
M$^3$AE\citep{chen2022multi} &  All& None & 0.746 & 0.571 & 0.410 & 0.576 \\
MedViLL\citep{moon2022multi}  &  All& None& 0.739 & 0.541 & 0.404 & 0.561 \\
Fusion Transformer &  All& None & 0.721 & 0.464 & 0.374 & 0.519 \\
Blind Transformer (Seeing questions only)  &  All& None& 0.677 & 0.310 & 0.240 & 0.409 \\
Deaf Transformer (Seeing ECGs only)  &  All& None & 0.673 & 0.314 & 0.270 & 0.419 \\
Med-Gemini\citep{saab2024capabilities} &  All & Tuning& -- & -- & -- & 0.584\\
\rowcolor{green!5}
\SysName (Ours)  &  All& Tuning & 0.909 & 0.603 & 0.329 & 0.614 \\

\bottomrule
\end{tabular}
\vspace{0.5cm}
\end{table*}

We present comprehensive benchmarking results comparing our proposed model, \SysName, with a variety of baseline approaches on the ECG-QA task, as shown in Table~\ref{tab:accuracy_comparison}. Overall, \SysName outperforms all existing baselines across both evaluation settings, using the full test set and the 10\% subset, achieving up to 4\% and 3\% higher average exact-match accuracy than the second-best models, respectively. This demonstrates \SysName’s robustness in both low-resource and full-data scenarios. When analyzing individual question types, \SysName consistently achieves the highest accuracy on both the verify and choose question types, which necessarily demand precise decision-making and fine-grained understanding. Notably, in the verify case under full test evaluation, \SysName reaches nearly 91\% accuracy, showing substantial gains over M$^3$AE (74.6\%) and ECG-ReGen (72.5\%). These improvements underscore the efficacy of our ECG-to-prefix alignment and the impact of dynamic prompting during proposed instruction tuning in enhancing the model’s capacity to process structured clinical queries. 

Regarding the query questions, which remain the most challenging for all models, primarily due to their open-ended nature with a diversity of answer choices, \SysName still performs competitively. Accordingly, in the full test set evaluation, while \SysName surpasses methods such as Blind Transformer, Deaf Transformer, and ECG-ReGen, it is slightly outperformed by approaches like M$^3$AE, MedViLL, and Fusion Transformer. However, it is important to emphasize that these competing models frame the task as a multi-answer classification problem~\citep{NEURIPS2023_d0b67349}, which reduces the complexity of the output space by relying on predefined label sets. In contrast, \SysName is designed for a more demanding generative modeling setting, where the model must produce free-form text responses conditioned on ECG and prompts. This setup requires more nuanced reasoning and sophisticated language generation, demonstrating the broader applicability and flexibility of our framework in real-world clinical question-answering scenarios. That being said, the performance gap in the query category is likely attributable to our use of a relatively compact and general-purpose model, LLaMA-Instruct 1B, which might lack prior exposure to biomedical texts or domain-specific fine-tuning. Alternatively, some counterpart models utilize larger or domain-specialized language backbones, such as BioMedGPT-LM-7B (in ECG-LM), which may provide stronger generative capabilities in complex ECG reasoning scenarios.

\subsubsection{Ablation Studies}
\label{subsub:ablation}

\begin{table}[t]
\scriptsize
\setlength\tabcolsep{3.5pt}
\centering
\caption{Impact of model components (report average of three question types). We evaluate the models on 10\% of testing data using the five metrics: EM-Acc, BERTScore, ROUGE, METEOR, and BLEU-1.}
\vspace{0.1cm}
\label{tab:ablation}
\begin{tabular}{@{}lccccc@{}}
\toprule
\textbf{Setting} & \textbf{EM-Acc} & \textbf{BERTScore} & \textbf{ROUGE} & \textbf{METEOR} & \textbf{BLEU-1} \\
\midrule
W/o DP     & 0.616 & 0.844 & 0.714 & 0.495 & 0.729 \\
W/o Retrieval Report       & 0.605 & 0.826 & 0.694 & 0.471 & 0.716 \\
W/o ET-Mapper    & 0.600 & 0.833 & 0.701 & 0.486 & 0.726 \\
W/o Pos-Encoder   & 0.582 & 0.815 & 0.669 & 0.461 & 0.696 \\
\midrule
Frozen LLM        & 0.583 & 0.820 & 0.677 & 0.453 & 0.694 \\
Frozen ECG Encoder & 0.587 & 0.829 & 0.687 & 0.481 & 0.715 \\
\midrule

\rowcolor{green!5}
\SysName (Ours)          & 0.628 & 0.845 & 0.727 & 0.502 & 0.744 \\
\bottomrule
\end{tabular}
\vspace{0.1cm}
\end{table}

To gain deeper insights into our \SysName model, we conduct ablation studies detailed in Table~\ref{tab:ablation}. These experiments systematically omit or change a specific component, allowing a determination of their impact on overall performance. Overall, we can see that each component of \SysName contributes meaningfully to its final result. For example, when removing dynamic prompting (DP), the performance slightly drops, approximately 1–2\% in EM-Acc and other metrics, while BERTScore remains relatively unchanged. This is expected, as techniques like option shuffling and random report selection contribute more to training robustness than to the overall semantic fidelity of the generated answers, which is what BERTScore primarily measures.

Following that, ablation results reveal more substantial performance degradation when removing key components related to modality alignment and clinical grounding. Specifically, removing the retrieved clinical report (W/o Retrieval Report) results in an average performance drop of approximately 2–3\% across most metrics, underscoring the importance of incorporating external domain-specific knowledge to enrich the model’s contextual understanding. Similarly, eliminating the ET-Mapper (W/o ET-Mapper), which is responsible for transforming ECG embeddings into LLM-compatible prefix representations, yields a comparable decline in performance. This confirms the critical role of this module in bridging the modality gap between ECG signal features and language token representations, facilitating effective multimodal fusion during decoding.

Notably, a pronounced degradation is also observed when omitting the positional layer (W/o Pos-Encoder), which removes the skip connection that conveys lead-specific information from the ECG encoder into the transformer mapper. This configuration leads to a drop of up to 4\% in EM-Acc and also substantial decreases in BERTScore (3\%) and other generative metrics such as ROUGE (6\%), METEOR (4\%), and BLEU-1 (5\%). The observed decline suggests that preserving positional leads, particularly in 12-lead ECGs, where each lead contributes unique spatial insights, is crucial for maintaining the semantic integrity of the generated answers. Without this lead-aware encoding, the model may struggle to differentiate among lead-specific features, potentially resulting in overfitting or entangled representations during prefix generation.

Similarly, we observe that freezing either the LLM or the ECG encoder leads to substantial performance degradation, with EM-Acc dropping by approximately 4\%, BERTScore decreasing by 2\%, and other generation-focused metrics such as BLEU-1, ROUGE, and METEOR reduced by 3–5\%. In particular, when the LLaMA-Instruct 1B model is frozen, it may struggle to adapt to the characteristics of ECG-derived inputs and the nuances of clinical language. Likewise, tuning the ECG encoder to adapt to the PTB-XL domain, especially when it was previously pretrained on a different dataset such as MIMIC IV ECG, also proves beneficial to the overall performance in the target ECG domain. 

\section{Discussion}
\SysName has demonstrated strong performance in the ECG-QA task, highlighting its potential to bridge physiological signals and language understanding for clinical applications. To further advance the development of ECG-based question-answering systems, we outline several future directions that emerge from our analysis: (1) First, we observe that questions requiring numerical reasoning, particularly within the query-type category, remain challenging. This is especially evident when accurate responses depend on precise physiological metrics from raw signals, such as P wave duration, QRS interval, or QT interval. This limitation arises from the fact that the ECG encoder is not explicitly optimized to extract such fine-grained clinical features, and LLMs alone lack the inductive bias to infer low-level signal semantics from high-dimensional embeddings. To overcome this, future work could incorporate conventional ECG-derived features (e.g., heart rate, heart rate variability, noise indicators, or various interval durations) into the prompt context~\citep{yu2023zero}, enhancing the model’s capacity for grounded clinical reasoning. (2) Second, our usage of the lightweight LLM due to resource constraints, while efficient, may limit generalization in biomedical domains due to its lack of medical pretraining. We believe that integrating larger, medically fine-tuned LLMs could further enhance answer reasoning capabilities in ECG-focused QA tasks. (3) Lastly, we highlight the potential of knowledge graph integration\citep{yan2024multi, hu2024interpretable}, where graph-structured clinical entities (e.g, Arrhythmia) and relations (e.g., Atrial Fibrillation shares some patterns with Atrial Flutter) are introduced to inject textual knowledge into ECG features, possibly improving question-answering performance and its interpretability. 
\section{Conclusion}
\label{sec:end}
We present \SysName, a novel multimodal LLM framework designed for ECG-based question answering. \SysName seamlessly integrates ECG signals with textual inputs through an instruction-tuned language model, supported by an ECG-aware transformer-based mapping module that aligns ECG embeddings with text representations. To enhance contextual reasoning and improve response quality, our approach leverages clinically relevant information retrieved from similar ECG cases and integrates it into dynamically customized prompts, tailored to each input. Through extensive evaluations on the benchmark dataset, \SysName achieves state-of-the-art performance in exact-match accuracy and shows strong competitiveness across generative evaluation metrics. These results highlight the model’s capability for nuanced clinical reasoning and emphasize its potential to facilitate more interpretable and effective ECG understanding in real-world healthcare applications.

\bibliography{references}

\end{document}